\documentclass[aps,prd,noshowpacs,preprintnumbers,nofootinbib,twoside]{revtex4} 

\usepackage{amsmath}
\usepackage{amssymb}
\usepackage{graphicx}
\usepackage{bm}
\usepackage{subfigure}

\newcommand{\Cal}[1]{\ensuremath{\mathcal{#1}}}

\newcommand{\eqn}[1]{Eqn. \eqref{#1}}
\newcommand{\eqns}[1]{Eqns. \eqref{#1}}
\newcommand{\Cite}[1]{Ref. \cite{#1}}
\newcommand{\Cites}[1]{Refs. \cite{#1}}

\newcommand{\fig}[1]{Fig. \ref{#1}}
\newcommand{\figs}[1]{Figs. \ref{#1}}

\newcommand{\ti}[1]{{\tilde #1}}
\newcommand{\wti}[1]{{\widetilde #1}}

\newcommand{\be}{\begin{equation}}
\newcommand{\ee}{\end{equation}}
\newcommand{\la}[1]{\label{#1}}

\newcommand{\p}{\ensuremath{\partial}}
\newcommand{\eps}{\ensuremath{\epsilon}}

\newcommand{\state}[1]{\ensuremath{|{\tt #1}\rangle}}

\newcommand{\avg}[1]{\ensuremath{\langle\,#1\,\rangle}} 

\newcommand{\UK}{\ensuremath{\Cal{U}}}
\newcommand{\VK}{\ensuremath{\Cal{V}}}

\newcommand{\Sch}{Schwarzschild}
\newcommand{\tort}{\ensuremath{x_\ast}}

\begin{document}

\title{Semiclassical environment of collapsing shells}
\author{Kinjal Banerjee}
\email{kinjal@iucaa.ernet.in}
\affiliation{IUCAA, Post Bag 4, Ganeshkhind, Pune - 411 007, India\\}

\author{Aseem Paranjape\footnote{Address from 8th
  Oct. 2009 : The Abdus Salam ICTP, Strada Costiera 11, 34014 Trieste,
  Italy.}}
\email{aparanja@ictp.it}
\affiliation{Tata Institute of Fundamental Research, Homi Bhabha
Road, Colaba, Mumbai - 400 005, India\\}

\date{\today}

\begin{abstract}
We explore in detail the semiclassical environment of collapsing
shells of matter, and determine the semiclassical flux measured by a
variety of observers. This study is a preliminary step in a broader
investigation of thermodynamic properties of the geometry of
collapsing objects. Specifically, in this paper we consider
spherically symmetric null and timelike collapsing shells which form
an event horizon, and calculate the flux measured by observers both
inside and outside the shell, and  both inside and outside the event
horizon, and find nontrivial results in most of the
cases. Additionally, we also investigate the environment of a shell
which collapses but \emph{does not} form a horizon, halting at some
radius larger than the \Sch\ radius, and find that such an object
generically gives rise to a pulse of radiation which is sharply peaked
as it travels inwards and is reflected at the origin, and eventually
emerges from the shell in a ``thermalized'' form. Our results have
potential consequences in addressing questions pertaining, e.g. to
black hole entropy and backreaction. 
\end{abstract}

\maketitle

\section{Introduction}
\noindent
The behaviour of quantum fields in the spacetime geometry of
black holes and gravitationally collapsing objects in general has been
a subject of detailed investigation since 
Hawking's seminal discovery \cite{hawking} of a nearly thermal flux of
particles that is measured by observers located outside the collapsing
object. Several authors have subsequently contributed towards building
an understanding of the semiclassical environment of collapsing
objects \cite{unruh,bir-dav,davies,semiclass-misc}. One of the most
fascinating aspects of this field of research is the fact that
classical black holes appear to behave like macroscopic
thermodynamical objects \cite{BHthermoD}. An important offshoot of
this behaviour is the so called information paradox (see,
e.g. \cite{mathur} for a recent interesting discussion), and an
understanding of this thermodynamics at the quantum level is also
likely to have consequences towards building a quantum theory of
gravity \cite{semiclass-QG}. 

There are several questions that can be asked within the
semiclassical framework. For example, considering a Schwarzschild
black hole for simplicity, it is a celebrated fact that the
black hole can be attributed an entropy which scales as one quarter of
the area of the event horizon in Planck units. This entropy however
has a physical relevance only to those observers who choose to
remain outside $r=2M$ \emph{forever}, i.e. -- for observers who 
truly experience $r=2M$ as an event horizon. Since forever is a
long time to wait, it becomes interesting to ask whether the concepts
of black hole entropy etc. can be meaningfully addressed
\emph{locally}. For example, what entropy (if any) would be attributed
to the surface $r=2M$ by observers who remain outside for an
arbitrarily long time, before suddenly choosing to jump into the black
hole? Such observers will clearly eventually have access to many
more microstates compared to their cousins who choose to remain
outside forever, but until the decision to fall in is actually made,
there are no grounds to distinguish between these two classes.

Returning to the context of collapsing objects, further complications
arise when one begins to account for the backreaction of the
semiclassical flux of particles, on the classical geometry of the
collapsing object. The basic idea here is that the loss of energy from
the collapsing object (which we take to be spherically symmetric, for
simplicity), in the form of the semiclassical flux, implies that the
correct background geometry in the exterior of the object 
must be time \emph{dependent}, rather than the static
\Sch\ spacetime \cite{brout,davies,semiclass-misc,backrxn}. Due to
technical challenges, the exact effect of the backreaction has been a
subject of much debate, with claims to the extent that the formation
of the event horizon itself may be hindered by the backreaction 
\cite{vachaspati,visser}. While this appears unlikely
\cite{paddy-aseem}, it is nevertheless true that determining the exact
geometry in the late stages of black hole \emph{evaporation}, remains
an open problem. Consequently, questions regarding the thermodynamic
behaviour of the geometry of the black hole/collapsing object also
remain clouded by the backreaction.  

In this paper, as a preliminary step towards addressing the broader
issues raised above, we set a simpler mandate for ourselves. We wish
to explore the \emph{full} semiclassical environment of a collapsing
object, which we will consider to be a spherically symmetric thin
shell \cite{unruh}. Classically, the exterior of the shell is
described by \Sch\ spacetime while the interior is Minkowski
spacetime. In particular we wish to determine the
semiclassical flux, as encoded in the renormalized stress tensor,
observed by a variety of observers, both inside and outside the shell,
before and after horizon formation.

There are several motivating factors behind such a study : Firstly,
while the flux seen by observers outside the shell and the event
horizon has been studied in great detail (beginning with Hawking's own
calculation), the nature of the flux inside the shell, and inside the
event horizon but outside the shell, remains largely unexplored. In
fact, an argument one usually encounters is that observers who fall
into a black hole do not encounter a significant flux. Our calculation
assumes zero backreaction, and we will see that in this case an
infalling observer will see a flux of outgoing particles which in fact
increases without bound as the observer approaches the singularity
(although as expected it is finite at all other locations, including
$r=2M$). Further, at least in the case of timelike collapse, we find that
observers \emph{inside} the shell also see a flux of particles, which
might have some interesting consequences which we will return to discuss
in the final section.

Apart from collapsing shells which form an event horizon, we will also
discuss the case, first described in \Cite{paddy-aseem}, of a shell
which even classically \emph{does not} form a horizon; instead it
asymptotically approaches a final radius larger than $2M$. In
\Cite{paddy-aseem} such a shell was shown to have very interesting
properties, although that discussion was limited partly to order of
magnitude estimates, since some required expressions were not
analytically obtainable. In this paper we will perform a full numerical
study of such a shell trajectory, and also explore its
semiclassical behaviour both inside and outside the shell. We
will find that such trajectories are generically expected to generate
a pulse of radiation which travels inwards, reflects at the center and
travels back outwards, eventually reaching an outside observer in a
``thermalized'' form (to be made precise later).

At this stage, it is worth addressing why we choose to study the
seemingly unrealistic situation of a collapsing shell. The first
reason of course is that the geometry of a shell is simple to handle
analytically, not least because the interior is just Minkowski
spacetime. A second important reason is that the absence of interior
matter allows us to study the purely gravitational aspects of quantum
fields in the interior. It is important to understand such effects
before addressing the problem of quantum fields whose semiclassical
effects interact with those of \emph{classical} matter inside a star
for example. Throughout our discussion we will simply prescribe the
classical trajectory of the shell, and not worry about the specific
(singular) stress-tensor needed to generate the trajectory. This is
not expected to interfere with the conclusions we draw regarding the
semiclassical behaviour of the shell.

The plan of the paper is as follows : In Section \ref{sec-null} we
start with the 
example of a \emph{null} shell which has been collapsing
forever. While such a geometry has been studied in the literature
\cite{davies,nullshell-1,nullshell}, we note that the region of
\Sch\ spacetime for $r<2M$ has been largely ignored. Another reason to
study the null 
shell is that all expressions for the flux, for all observers, can be
obtained analytically, with no numerical inversions involved. This
section will therefore serve as a useful warmup to the more involved
calculations for timelike trajectories. In Section \ref{sec-time} we
turn to studying timelike trajectories, where many of the calculations
will be numerical. We will first study a trajectory which forms a
horizon, describing the flux seen by various observers, and then turn
to the trajectory mentioned earlier, which asymptotically halts
without forming a horizon. We will end in Section \ref{discuss} with
a brief discussion and prospects for future work.

\subsection{Notation} 
\la{sec-notation}
\noindent
To avoid confusion, in this subsection we will lay down all the
notation that will be used subsequently and in the next subsection we
recapitulate the prescription for calculating the semiclassical flux
in a given geometry. Note that all our calculations assume zero
backreaction.  

We will be assuming Planck units throughout, with $G$,
$\hbar$ and $c$ set to unity. We will mostly follow the notation used
in \Cite{paddy-aseem}, which in turn was based on the notation in the
review by Brout {\it et al.} \cite{brout}. Additionally, to avoid
cumbersome factors of $2M$ (where $M$ is the mass of the collapsing
object), we will assume that all the coordinates are
non-dimensionalized by dividing out by $2M$. This will lead to
expressions for the flux, etc. with factors of $2M$ missing, which can
be put back appropriately by dimensional arguments. Latin tensor
indices $a,b,..$ range over $0..3$, and we use the mostly plus
sign convention.   

As mentioned earlier, the exterior of the shell is \Sch\ spacetime
while the interior is Minkowski, and we can use the same radial
coordinate $r$ to foliate both regions. We refer to \Sch\ time as $t$
and Minkowski time as $T$. Outgoing null coordinates
will be a variation of the letter $u$, similarly ingoing coordinates
and $v$. To be precise, the interior Minkowski null coordinates will
be
\be
V = T + r ~~;~~ U = T - r\,,
\la{not-1}
\ee
and the exterior Eddington null coordinates will be
\be
v = t + \tort(r) ~~;~~ u = t - \tort(r)\,,
\la{not-2}
\ee
where we have defined the tortoise function
\be
\tort(r) = r + \ln|r-1| \,,
\la{not-3}
\ee
(which should be read as $\tort(r)/2M = r/2M + \ln|r/2M-1|$). All
``Kruskalized'' (i.e. -- exponentiated) versions of the null
coordinates will be referred to using calligraphic
fonts. Specifically, \UK\ and \VK\ will denote the standard 
Kruskal null coordinates
\be
\VK = 2e^{v/2} ~~;~~ \UK = \mp 2e^{-u/2} ~~;~~ \UK\VK= 4e^r(1-r)\,, 
\la{not-4}
\ee
with the upper sign in \UK\ for $r>1$ (the \Sch\ radius being unity in
non-dimensionalized form) and the lower sign for $r<1$. Other
Kruskalized coordinates will be prescribed as needed.

With this setup, the (non-dimensionalized) metric in various regions
and in various coordinates becomes (suppressing the line element for
the $2$-sphere, which is always $r^2d\Omega^2$),
\begin{align}
ds^2_{\rm int} &= -dT^2 + dr^2 = -dUdV\,,\nonumber\\
&\nonumber\\
ds^2_{\rm ext} &= -(1-1/r)dt^2 + (1-1/r)^{-1}dr^2 \nonumber\\
&= -\left(1-\frac{1}{r}\right)dudv = -\frac{e^{-r}}{r}d\UK d\VK\,.
\la{eq-metric}
\end{align}
The subscript $s$ will always refer to evaluation on the shell, so the
trajectory is $r=R_s$, $v=v_s$, etc. The specific parameter used along
the trajectory will be chosen according to convenience, and could be
one of the coordinates.

\subsection{Recap of semiclassical prescription}
\la{sec-recap}
\noindent
Throughout, we will be interested in the semiclassical flux $\Cal{F} =
n_a\avg{T{}^a_b}u^b$ measured by an observer with 4-velocity $u^a$
where $n_au^a=0$, $n_an^a=1=-u_au^a$, and $\avg{T_{ab}}$ is the
renormalized stress-tensor vacuum expectation value (VEV) of a
quantized, massless scalar field in a specific vacuum state. We refer
the reader to \Cites{bir-dav,brout} for detailed treatments of the
formalism, and to \Cite{paddy-aseem} for a pedagogical
introduction. The Hawking flux has the value $(\pi/12)T_H^2 =
(192\pi)^{-1}(2M)^{-2}$, where $T_H=1/8\pi M$ is the Hawking 
temperature. The vacuum state \state{in} to be used, is completely
defined by specifying its behaviour at past null infinity $\Cal{I}^-$
and by requiring that its mode functions vanish at reflection at the
origin $r=0$. Typically one requires that the vacuum on $\Cal{I}^-$ be 
identical to the Minkowski vacuum, so that the ingoing modes of the
state \state{in} become $\sim e^{-i\omega v}$ for a typical timelike
collapse. We will see that the situation for a null shell collapsing
forever, is somewhat different, since in this case the portion of
$\Cal{I}^-$ relevant for thermal flux in the exterior, lies
\emph{inside} the shell. However, here also we will be able to
meaningfully compute the flux observed by specific classes of
observers. 

An important assumption we make is to work in the $1+1$ $s$-wave
approximation, ignoring the \Sch\ potential barrier, so that the
results of $2$-dimensional conformal field theory (CFT) apply. For a
discussion of this approximation and its validity, see the review by
Brout {\it et al.} \cite{brout}. In this approximation, for a vacuum
state \state{vac} with ingoing and outgoing modes given respectively by 
$e^{-i\omega f_-}$ and $e^{-i\omega f_+}$, such that the metric in the
relevant region in terms of these null functions is $ds^2 =
-C(f_+,f_-)df_-df_+$, the components of the renormalized stress-tensor
VEV in the state \state{vac} are calculated as
\be
\avg{T_{f_-f_-}}_{\rm vac} =  -F_{f_-}[C(f_-,f_+)] ~~;~~
\avg{T_{f_+f_+}}_{\rm vac} = -F_{f_+}[C(f_-,f_+)] \,,
\la{eq-flux}
\ee
where the derivative operator $F$ is defined, for any function $f(x)$,
as
\be
F_x\left[f(x)\right] \equiv \frac{1}{12\pi} f^{1/2}\p^2_xf^{-1/2} =
-\frac{1}{48\pi}\left[ 2\p^2_x\ln f - \left(\p_x\ln f\right)^2
  \right]\,. 
\la{eq-deriv1}
\ee
The cross terms $\avg{T_{f_-f_+}}_{\rm vac}$ can be determined by
requiring that the stress tensor be covariantly conserved. The
following relations are useful while analyzing the flux in different
regions : 
\begin{subequations}
\begin{align}
F_x[x] &= \frac{1}{16\pi x^2} \,,
\la{eq-deriv2a}\\
F_x[1/f(x)] &= -F_x[f(x)] + \frac{1}{24\pi}(\p_x\ln f)^2\,,
\la{eq-deriv2b}\\
F_x[f(x)g(x)] &= F_x[f(x)] + F_x[g(x)] + \frac{1}{24\pi}(\p_x\ln
f)(\p_x\ln g)\,,
\la{eq-deriv2c}\\
F_s[f(x(s))] &= \frac{1}{s^{\prime2}}\left( F_x[f(x)] + \frac{1}{24\pi}
(\p_x\ln f)(\p_x\ln s^\prime)\right)\,,
\la{eq-deriv2d}
\end{align}
\la{eq-deriv2}
\end{subequations}
where in the last equation, $s^\prime=ds/dx$.

Hawking's result follows from recognizing that for a generic timelike
collapse, requiring the vacuum \state{in} to be Minkowskian at
$\Cal{I}^-$, the corresponding modes in the exterior region become
$e^{-i\omega v}$, $e^{-i\omega G(u)}$, where $G(u)$ is that value
$v=G$ for which an ingoing ray of constant $G$, becomes a ray of
constant $u$ after being reflected at the origin and emerging from the
shell. Further, one finds that due to the presence of the event
horizon, the generic behaviour of $G(u)$ for $u\to\infty$ is
$dG/du\sim e^{-u/2}$, which leads to the result that in the state
\state{in}, an exterior observer using the Eddington $u,v$ coordinates
measures a constant outgoing thermal flux $\avg{T_{uu}}-\avg{T_{vv}} = 
(\pi/12)T_H^2$ as $u\to\infty$. [The notation $\avg{T_{ab}}$ with no
  subscript will always refer to the vacuum state \state{in}.] We will 
see this result in the trajectories we study. For further details see
\Cites{bir-dav,brout,paddy-aseem}.

\section{Null shell}
\la{sec-null}
\begin{figure}[t]
\begin{center}
\includegraphics[height=0.225\textheight]{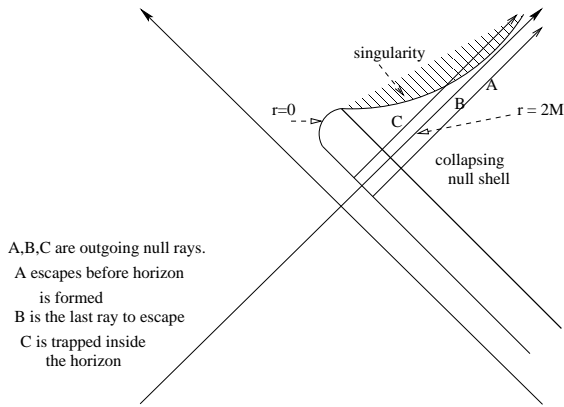}
\end{center}
\caption{Penrose diagram for the null shell that has been collapsing
  for all time \cite{nullshell-1}. Ingoing rays have $\VK=\,$const and
  outgoing rays have $\UK=\,$const.}
\la{penrose-null}
\end{figure}
\noindent
We begin by considering a null shell that has been collapsing for all
time \cite{davies,nullshell-1,nullshell}. The global structure of this
system is different from the usual timelike collapse situations which
we will discuss in Section \ref{sec-time}, since a part of $\Cal{I}^-$
in this case lies \emph{inside} the shell. Since the matching of
metrics in \eqref{eq-metric} occurs along a line of constant $V$
(which we conveniently choose to be $V=0$), there is a freedom in
relating the exterior and interior ingoing coordinates. We exploit
this freedom to set $\Cal{V}_{\rm m} \equiv 2e^{V/2} =
\VK$. Additionally, we use the Kruskalized outgoing coordinate
$\Cal{U}_{\rm m} \equiv 2e^{-U/2}(1+U/2)$, and it is then easy to show
that on the shell (i.e. -- at $\Cal{V}_{\rm m} = \VK=2$) we have
$\Cal{U}_{\rm m}=\UK$ for all $r$. In other words, we can use the
usual Kruskal coordinates \UK, \VK\ over the entire spacetime, with
the metric given by 
\begin{align}
ds^2_{\rm int} &= -dUdV = -\frac{2e^{-r}}{(-U)}d\UK d\VK
\,,\nonumber\\ 
ds^2_{\rm ext} &= -\frac{e^{-r}}{r}d\UK d\VK\,,
\la{eq-null-1}
\end{align}
with the event horizon $r=1$ corresponding to the outgoing ray $\UK=0$
or $U=-2$. Figure \ref{penrose-null} shows the Penrose diagram for
this spacetime \cite{nullshell-1}. 

In the usual timelike collapse situations where the shell would
typically start collapsing from some finite radius, $\Cal{I}^-$ lies
entirely outside the shell and the spacetime is asymptotically
flat. It is then natural to choose a vacuum whose ingoing modes at
$\Cal{I}^-$ are $\sim e^{-i\omega v}$, $v$ being the ingoing Eddington
coordinate. In the present situation, it is natural to define the
vacuum in the interior region of $\Cal{I}^-$ using the
\emph{Minkowski} ingoing coordinate $V$. Note however that our choice
of mapping the ingoing coordinates between the interior and exterior,
ensures that $V=v$ at the boundary, and hence the \state{in} vacuum can still be
chosen to have modes $\sim e^{-i\omega v} = e^{-i\omega V}$ on
\emph{all} of $\Cal{I}^-$. The reflection condition at $r=0$ then
fixes the outgoing modes of \state{in} to be $\sim e^{-i\omega  U}$ in the interior 
region. Equivalently, the function $G$ discussed in Section
\ref{sec-recap} is simply $G(U)=U$ in the interior. It then follows
that the renormalized stress tensor in the state \state{in} is
identically zero in the interior of the shell, since the conformal
factor in the interior is constant in the $U$,$V$ coordinates. 

In the exterior, for $r>1$ we want the flux measured by static
observers $\Cal{O}_1$ who use the Eddington $u$,$v$ coordinates. The
function $G(u)$ is now determined by the implicit equation
\be
-e^{-u/2} = e^{-G/2}\left(1+G/2\right)\,.
\la{eq-null-2}
\ee
The stress tensor components in the state \state{in} are given by
\be
\avg{T_{vv}}=
-\left(\frac{\pi}{12}T_H^2 \right)
\frac{1}{r^3}\left(4-\frac{3}{r}\right) ~~;~~  
\avg{T_{uu}}= \avg{T_{vv}} + F_u\left[ \frac{dG}{du} \right]\,. 
\la{eq-null-3}
\ee
[Note that the $r$-dependent part of these expressions is simply the
static contribution in the so called Boulware state.] The outgoing
flux measured by observers $\Cal{O}_1$ is 
\begin{align}
\Cal{F} = -\avg{T{}^r_t} &= \avg{T_{uu}} - \avg{T_{vv}} \nonumber\\
&= F_u\left[ \frac{dG}{du} \right] = 
-\frac{1}{12\pi}\frac{1}{G(u)^4}\left(3+2G(u)\right)\,. 
\la{eq-null-4}
\end{align}
We can analytically see that as $u\to\infty$, $G\to-2$, $dG/du \sim 
e^{-u/2}$, and hence the flux asymptotically approaches the Hawking
value (note that in our convention for non-dimensionalization, the
Hawking value is $1/192\pi$). Figure \ref{nullflux} shows this
flux (normalized by the Hawking value) as a function of $u$. 
\begin{figure}[t]
\begin{center}
\includegraphics[height=0.2\textheight]{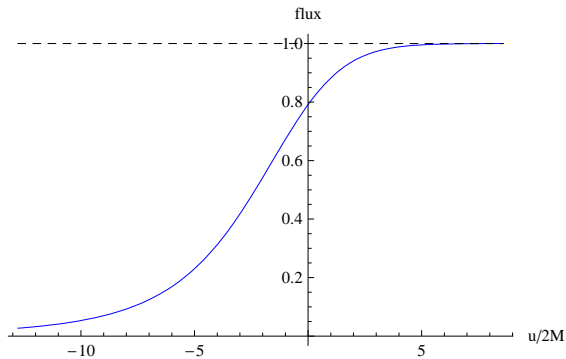}
\end{center}
\caption{The semiclassical flux (normalized by the Hawking value)
  measured by static observers outside the event horizon, for the null
  shell trajectory.}
\la{nullflux}
\end{figure}

An interesting region to consider is the exterior of the shell for
$r<1$. In this region it is more natural to use the Kruskal
coordinates \UK,\VK, since this region is likely to be accessed by
freely falling observers $\Cal{O}_2$. The stress tensor (in the entire
exterior region $r>R_s$) in these coordinates is
\begin{align}
\avg{T_{\VK\VK}} &= \frac{4}{\VK^2} \avg{T_{vv}}\,,\nonumber\\  
\avg{T_{\UK\UK}} &= \frac{4}{\UK^2} \avg{T_{uu}}\,,
\la{eq-null-5}
\end{align}
Although superficially divergent, the expression for
$\avg{T_{\UK\UK}}$ is actually finite on the horizon $\UK=0$. The
structure of $\avg{T_{vv}}$ and $\avg{T_{uu}}$ shows, on the other
hand, that observers who fall into the singularity will generically
see a flux that increases without bound. We will see this feature
again when studying timelike trajectories in Section \ref{sec-time}
below, where we will display a more detailed expression and a plot for 
the flux seen by observers $\Cal{O}_2$. A third class $\Cal{O}_3$
which remains in free fall \emph{inside} the shell for a certain time, does
not see a flux in the present null case. The timelike case however
will prove to be different. 

We end this section by noting that although inertial observers in the
interior of the null shell do not see a semiclassical flux, there is a
natural class of accelerated observers who \emph{do} see a flux. These are 
observers following trajectories picked out by the \emph{Kruskal} null coordinates, 
and are related to Rindler observers in the Minkowski region. (Notice the contrast with 
the exterior, where such Kruskal observers are \emph{inertial}, while observers following
trajectories picked out by \emph{Eddington} coordinates are
Rindler-like.) For a vacuum state \state{vac} defined by the modes $e^{-i\omega
  \VK}$, $e^{-i\omega\UK}$ in the interior, the stress-tensor
components are given by
\begin{align}
\avg{T_{\VK\VK}}_{\rm vac} &= \left(\frac{\pi}{12}T_H^2 \right) 
e^{-V}\,, \nonumber\\ 
\avg{T_{\UK\UK}}_{\rm vac} &= \left(\frac{\pi}{12}T_H^2 \right) 
 \frac{4e^U}{U^4} \left( U^2 - 4U + 12\right)\,.
\la{eq-null-6}
\end{align}

\section{Timelike shell}
\la{sec-time}
\noindent
We now turn to the more realistic situation of a shell collapsing
along a timelike trajectory. We will discuss two types of trajectories
: the first is a ``conventional'' trajectory in which the shell forms
a horizon in finite proper time, and continues to collapse to a
singularity. The second is an ``asymptotic'' trajectory, in which the
shell asymptotically approaches a final radius \emph{larger} than its
\Sch\ radius, so that a horizon is never formed. In \Cite{paddy-aseem}
the latter was shown to have very interesting behaviour, in
that an outside observer would see a constant Hawking-like flux for
some finite time interval, after which the flux would exponentially
decay to zero. In \Cite{paddy-aseem} however, only a part of the flux
was explicitly calculated, with order of magnitude estimates for the
remaining part, since the functional inversions needed could not be
performed analytically. Below we will show the results of a
full numerical calculation of the flux, which will corroborate the
results of \Cite{paddy-aseem}. We will also see some
additional effects as mentioned in the introduction, which are only
revealed in the numerical analysis. All numerical calculations were
performed on \emph{Mathematica}. We begin with a discussion of the
conventional horizon-forming trajectory.

\subsection{Horizon-forming trajectory}
\noindent
We parametrize the trajectory using the shell proper time. Our
trajectory is $r=R_s(\tau)$ (assuming a non-dimensionalised form for
all coordinates), where
\be
R_s^{\prime 2} = \theta(\tau)\frac{f(\tau)}{R_s} ~~;~~ R_s(\tau) =
R_0,~~ \tau\leq0\,,
\la{eq-shell}
\ee
where a prime denotes a derivative with respect to the argument
$\tau$, $\theta$ is the Heaviside step function, and $f$ is a function
which is chosen so that $f\to1$ at late times. Note that
$R_s^\prime\leq0$. The trajectory therefore remains at a constant
radius $R_0$ until $\tau=0$, and eventually becomes a pseudo-free-fall
curve, mimicking a particle freely falling in the potential of a body
of mass $M$. The specific form of $f$ we choose is
\be
f(\tau) = 1 - e^{-(\kappa\tau)^a}\,,
\la{eq-foftau}
\ee
with $\kappa$ and $a$ as parameters. We will fix $a=5$, which ensures
that $R_s(\tau)$ has a continuous third derivative at $\tau=0$ (and
hence for all $\tau < \tau_{\rm max}$, where $R_s(\tau_{\rm
  max})=0$). Since the parametrization is in terms of the shell's
proper time, the equations \eqref{eq-metric} are sufficient to
completely determine the trajectory in terms of other coordinates
$u$, $v$, etc. as well.  In
particular, the functions $v_s$, $U_s$ are
useful in the analysis, and can be obtained by integrating the
following expressions 
\begin{align}
v_s^\prime &= 1/g ~~;~~ g(\tau) \equiv -R_s^\prime +
\sqrt{R_s^{\prime2}+1-1/R_s}\,,\nonumber\\
U_s^\prime &= -R_s^\prime+\sqrt{R_s^{\prime2}+1}\,,
\la{eq-traj}
\end{align}
and when needed, we have $V_s = U_s + 2R_s$, $u_s = v_s -
2x_\ast(R_s)$.  

As in the null case, we will calculate the semiclassical flux in this
spacetime, for three classes of observers $\Cal{O}_1$, $\Cal{O}_2$ and 
$\Cal{O}_3$. Observers $\Cal{O}_1$ and $\Cal{O}_2$ always remain
outside the shell, while observers $\Cal{O}_3$ remain inside the shell
until some time. More precisely, $\Cal{O}_1$ correspond to
$r=\,$const observers with $r>R_0$, $\Cal{O}_2$ are free-fall
observers who cross the event horizon and fall into the singularity,
and $\Cal{O}_3$ are chosen to be $r=\,$const observers \emph{inside}
the shell, who are tracked until the time when the shell crosses their
location. Figure \ref{penrose-hor} shows the Penrose diagram for this
spacetime, with example trajectories of $\Cal{O}_1$, $\Cal{O}_2$ and
$\Cal{O}_3$ observers also shown.
\begin{figure}[t]
\centering
\includegraphics[height=0.25\textheight,angle=45]{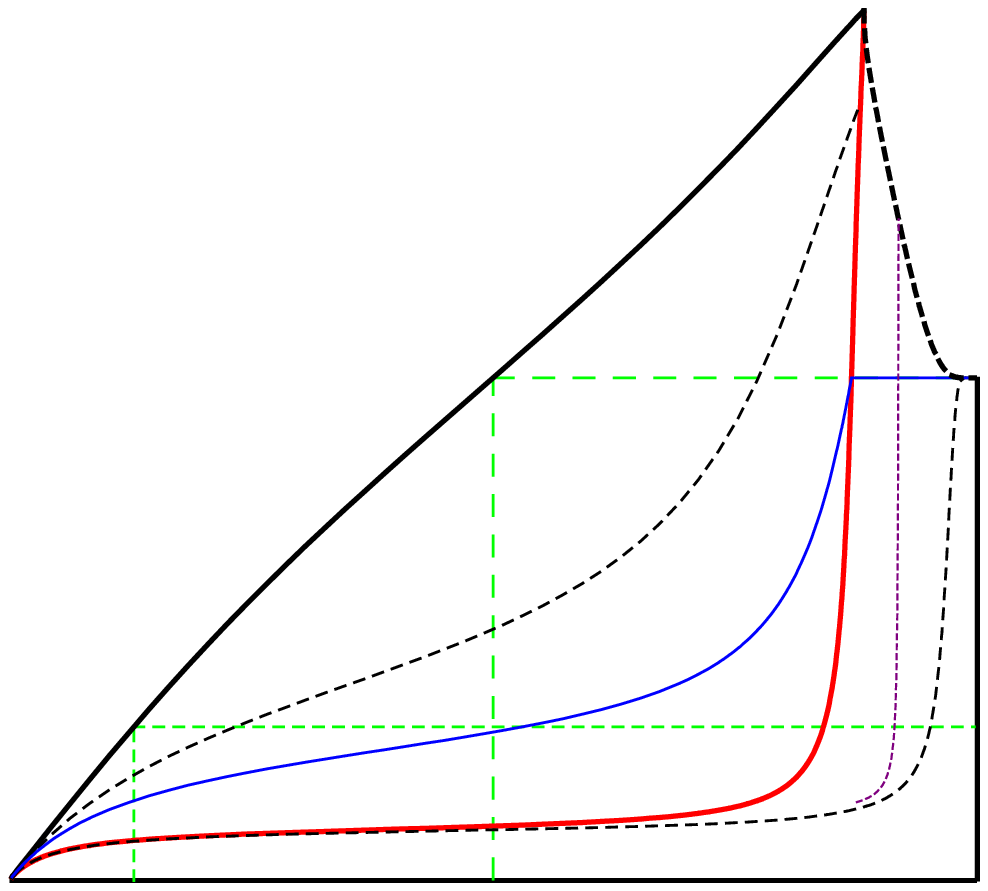}
\caption{\small Penrose diagram for timelike collapse of a shell, with 
  $R_0=6M$, $2M\kappa = 0.3$. The origin (leftmost thick black curve)
  remains regular until the shell collapses to the singularity (thick
  dashed black). Various lines correspond to the shell trajectory (thick
  solid red), the surface $r=2M$ (thin solid blue), the last escaper
  null ray (long dashed green), the ray which enters at $\tau=0$
  (short dashed green), and example trajectories of observers
  $\Cal{O}_1$ ($r=6.15M$, exterior short dashed black), $\Cal{O}_2$
  ($\beta=-0.5$, $\gamma=40.4M$, fine dashed purple) and $\Cal{O}_3$
  ($r=M$, interior short dashed black).}
\la{penrose-hor}
\end{figure}
Recall from Section \ref{sec-recap} that the modes corresponding to
the vacuum state \state{in} for this spacetime, are completely fixed
by requiring that (a) the modes on $\Cal{I}^-$ become $\sim
e^{-i\omega v}$, and (b) the modes vanish at the origin $r=0$. Given
these conditions, we can calculate the function $G$, which
then determines the form of the conformal factor used to calculate the
flux in \eqn{eq-flux}. 

\subsubsection{Flux for observers $\Cal{O}_1$}
\noindent
The calculation in this subsection is well known in the literature
(see, e.g. \Cite{bir-dav}). We will nevertheless display some relevant
details which demonstrate the general procedure followed in the
subsequent calculations. For observers $\Cal{O}_1$ with constant $r$
who use the Eddington $u$ as their natural retarded time, the function
$G(u)$ can be parametrized by $\tau$, with $u=u_s(\tau)$, and is given
by 
\be
G(\tau) = v_s(\tau_1(\tau)) ~~;~~ V_s(\tau_1) = U_s(\tau)\,,
\la{eq-Gob1-1}
\ee
i.e., a ray which exits the shell at $\tau$, must have entered at a
time $\tau_1$ determined by the second equation in \eqref{eq-Gob1-1},
which is just the reflection condition. $G$ is then by definition the
value of $v$ at the entry point $\tau_1$, and is treated as a function
of $\tau$. Further note that
\be
\frac{dG}{du} = \frac{dv_s(\tau_1)}{d\tau_1}\frac{d\tau_1}{d\tau}
\frac{1}{(du_s/d\tau)} = \tau_1^\prime v_s^\prime(\tau_1)/u_s^\prime\,,
\la{eq-Gob1-2}
\ee
where we have introduced the convenient shorthand notation of using a
prime to denote a derivative with respect to the argument, and
evaluation of all functions will be at their natural argument $\tau$,
unless explicitly stated as in the case of $v_s^\prime(\tau_1)$. We
will use this type of analysis for the observers $\Cal{O}_2$ and
$\Cal{O}_3$ as well. 

The (outgoing) flux \Cal{F}\ for these observers is independent of
$r$, and is given by (see also \Cite{bir-dav})
\be
\Cal{F}(u) = -\avg{T{}^r_t} = \avg{T_{uu}} - \avg{T_{vv}} = F_u\left[ 
  \frac{dG}{du} \right]\,,
\la{eq-fluxob1-1}
\ee
where the derivative operator on the extreme right was defined in
\eqn{eq-deriv1}. Using the relations in \eqns{eq-deriv2}, and
parametrizing again using $\tau$, we can rewrite this as
\be
\Cal{F}(\tau) = \frac{1}{u_s^{\prime2}} \bigg\{
-F_\tau[u_s^\prime(\tau)] +
\tau_1^{\prime2}(F_y[v_s^\prime(y)])|_{y=\tau_1(\tau)} +
F_\tau[\tau_1^\prime(\tau)] \bigg\} \,.
\la{eq-fluxob1-2}
\ee
The expression on the right can be evaluated numerically, with the
only nontrivial part being the inversion involved in determining
$\tau_1(\tau)$. The result is shown in \fig{flux-ob1}, where the flux
is plotted as a function of $u$. As expected, this flux
attains a constant value $(\pi/12)T_H^2$ at late times.
\begin{figure}[t]
\centering
\includegraphics[height=0.2\textheight]{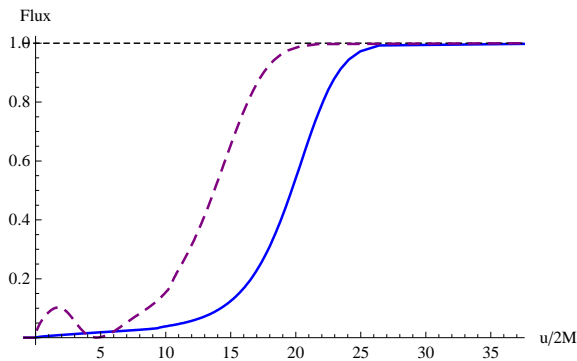}
\caption{\small The flux (normalized by the Hawking value
  $(\pi/12)T_H^2$) as a function of $u$, for the timelike
  horizon-forming trajectory with $R_0=8M$, for $r=\,$const observers
  outside the shell. The two curves correspond
  to $2M \kappa=0.1$ (solid blue) and $2M \kappa=0.3$ (dashed
  purple). The initial behaviour of the flux depends on the trajectory
  details, but eventually the flux attains the constant Hawking value
  (short dashed black line). }
\la{flux-ob1}
\end{figure}

\subsubsection{Flux for observers $\Cal{O}_2$}
\noindent
We take observers $\Cal{O}_2$ to be freely falling into the horizon
and eventually into the singularity. Defining the timelike and
spacelike Kruskal coordinates
\be
t_{\rm K} = \frac{1}{2}(\VK+\UK) ~~;~~ r_{\rm K} =
\frac{1}{2}(\VK-\UK)\,, 
\la{eq-coordK}
\ee
such an observer can be described as moving at constant velocity in
the $(t_{\rm K},r_{\rm K})$ coordinates, so that
\be
r_{\rm K} = \beta t_{\rm K} + \gamma/2\,,
\la{eq-trajob2-1}
\ee
for some constant $\gamma$ and $\beta<1$. This trajectory can also be
parametrized using the \emph{shell's} proper time $\tau$, and we
ensure that this observer always remains outside the shell by choosing
an appropriate $\gamma$ and tracking the trajectory only after a fixed 
time $\tau_{\rm ob,min}$. (Recall that for the eternal \Sch\ black
hole, a trajectory such as \eqn{eq-trajob2-1} corresponds to an
observer who pops out of a white hole at $t=-\infty$, reaches a
maximum radius $r=r_{\rm max}$, and falls into the horizon at
$t=+\infty$.) It is convenient to write the trajectory in terms of the
$r$ and \VK\ coordinates, as $r=R_{\rm ob}(\tau)$, $\VK=\Cal{V}_{\rm 
  ob}(\tau)$, where
\begin{align}
\Cal{V}_{\rm ob}(\tau) &= \left((1+\beta)\Cal{U}_s + \gamma
\right)/(1-\beta)\,,\nonumber\\ 
4e^{R_{\rm ob}}(1-R_{\rm ob}) &= \Cal{U}_s\Cal{V}_{\rm ob} \,,
\la{eq-trajob2-2}
\end{align}
where we have used the relation between $r$ and the Kruskal
\UK,\VK\ given by $\UK\VK=4e^r(1-r)$, and $\Cal{U}_s(\tau)$ is the 
trajectory equation which can be obtained from the matching conditions
on the shell,
\be
\Cal{U}_s(\tau) = 2e^{R_s-v_s/2}(1-R_s) \,.
\la{eq-trajob2-3}
\ee
The outgoing flux \Cal{F}\ for such an observer becomes 
\be
\Cal{F}=\left.n_a\avg{T{}^a_b}u^b\right|_{\rm ob} =
-\left.\left(re^{r}\left\{\avg{T_{\VK\VK}} - \avg{T_{\UK\UK}}
\right\}\right)\right|_{\rm ob}\,,
\la{eq-fluxob2-1}
\ee
so that the flux normalized by the Hawking value, denoted
$\Cal{\wti{F}}$, is 
\begin{align}
\Cal{\wti{F}} &=
-4\left.\bigg[\,\frac{e^r}{\VK^2 r^2}\left(\frac{3}{r}
  - 4 \right) - \frac{\VK^2}{16r}e^{-r} \left(1 + 
  \frac{2}{r} + \frac{3}{r^2}\right)
  \bigg]\right|_{\rm ob} 
+ 192\pi\left(re^{r} F_{\UK}\left.\left[
  \frac{dG}{d\UK}\right]\right)\right|_{\rm ob}\,.
\la{eq-fluxob2-2}
\end{align}
Note that while it seems analytically easier to parametrize this
trajectory using the Kruskal coordinate \Cal{U}, numerically it is
more convenient to use the shell proper time, due to the appearance of
the function $G$ in the flux. Since we are parametrizing the
observer's trajectory using the shell's proper time, $r$ and $\VK$ in 
\eqn{eq-fluxob2-2} will be replaced by $R_{\rm ob}(\tau)$ and
$\Cal{V}_{\rm ob}(\tau)$ respectively, and the last term involving the 
$F$ operator has a form identical to the RHS of \eqn{eq-fluxob1-2},
with $u_s(\tau)$ replaced with $\Cal{U}_s(\tau)$.

This flux is also straightforward to compute numerically, and the
results are shown in \fig{flux-ob2}, for a shell trajectory with
$R_0=10M$ and $2M\kappa=0.1$, for an observer with $\beta=-0.75$ and
$\gamma=27.7M$. We see that while the flux is small (with magnitude
comparable to the Hawking flux) as long as the
observer remains outside the horizon, and remains finite at horizon
crossing as well, once the observer approaches the singularity, the
flux increases without bound. This is also obvious from the expression
\eqref{eq-fluxob2-2}, where the first term in square brackets is
singular as $r\to0$. 
\begin{figure}[t]
\centering
\includegraphics[height=0.2\textheight]{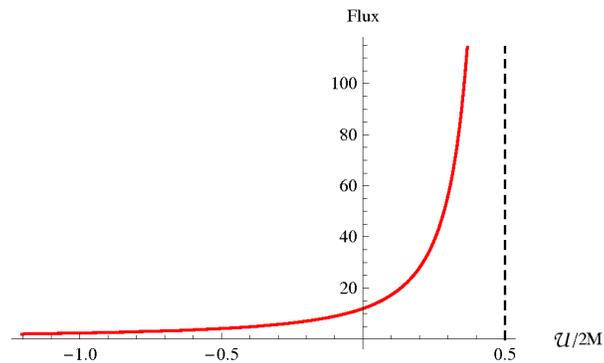}
\caption{\small The normalized flux (red solid line) for the
  timelike horizon-forming trajectory with $R_0=10M$ and
  $2M\kappa=0.1$, for a freely falling observer with $\beta=-0.75$ and 
  $\gamma = 27.7M$. The flux is plotted as a function of retarded
  Kruskal time \UK, with horizon formation occurring at $\UK=0$. The
  vertical dashed line indicates the time at which this observer hits
  the singularity}
\la{flux-ob2}
\end{figure}

\subsubsection{Flux for observers $\Cal{O}_3$}
\noindent
We now turn to the interior Minkowski region, and consider the flux
measured by $r=\,$const observers during the time that they remain
inside the shell. We do not track such observers once the shell has
crossed them, since this event is expected to be singular for such
observers even classically, and we have not been accounting for the
classical stress-tensor of the shell in our analysis. Unlike the null
case, here we will see that interior observers do detect a nonzero
flux, which is a consequence of the difference in global structure of
$\Cal{I}^-$ in the two cases.

The $r=\,$const observers inside the shell use $T=(V+U)/2$ as their
natural proper time. For any event $\Cal{P}=(U_\ast,V_\ast)$ along the
trajectory of such an observer, there are two events $\Cal{P}_1$ and
$\Cal{P}_2$ relevant for the flux calculation. $\Cal{P}_1$ is defined
as the entry point (into the shell) of a ray which, after reflection
at the origin, becomes the outgoing ray $U=U_\ast$. $\Cal{P}_2$ is
similarly the entry point of the ingoing ray which after entering the
shell becomes $V=V_\ast$. Let $\tau_1$ and $\tau_2$ be the values of
the shell proper time corresponding to the events $\Cal{P}_1$ and
$\Cal{P}_2$ respectively. In terms of $(T,r)$ we have
\be
V_s(\tau_1) = U_\ast= T-r ~~;~~ V_s(\tau_2) = V_\ast = T+r\,.
\la{eq-fluxob3-1}
\ee
The modes of the \state{in} vacuum are then
given by $\sim e^{-i\omega v_s(\tau_2)} - e^{-i\omega v_s(\tau_1)}$. 
Using the CFT rules for calculating the flux, it is then not hard to
show that the normalized outgoing flux
$\Cal{\wti{F}}=(\avg{T_{UU}}-\avg{T_{VV}})/(\pi T_H^2/12)$ for such
observers is given by  
\be
\Cal{\wti{F}} = \Cal{S}(\tau_1) - \Cal{S}(\tau_2)\,,
\la{eq-fluxob3-2}
\ee
where the function $\Cal{S}(\tau)$ is defined as
\be
\Cal{S}(\tau) = \frac{192\pi}{V_s^{\prime2}(\tau)}\left(
F_\tau[v_s^\prime(\tau)] - F_\tau[V_s^\prime(\tau)] \right)\,.
\la{eq-fluxob3-3}
\ee
\begin{figure}[t]
\centering
\subfigure[]{
\includegraphics[width=0.45\textwidth]{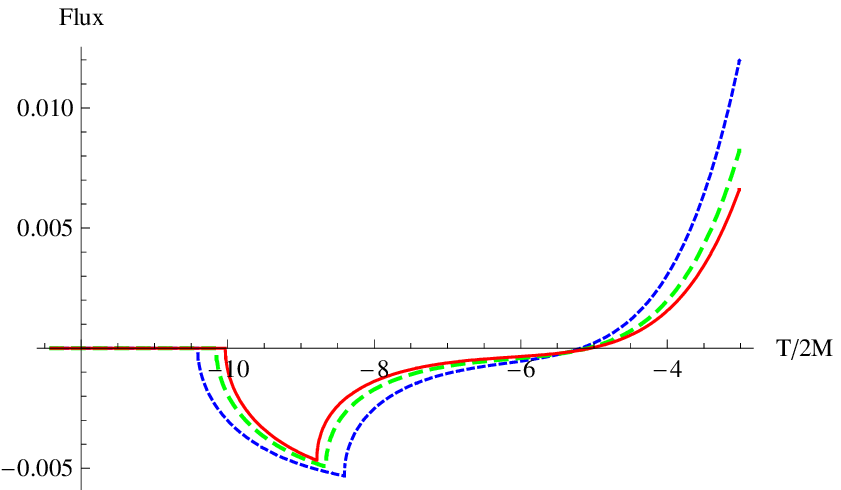}}
\subfigure[]{
\includegraphics[width=0.45\textwidth]{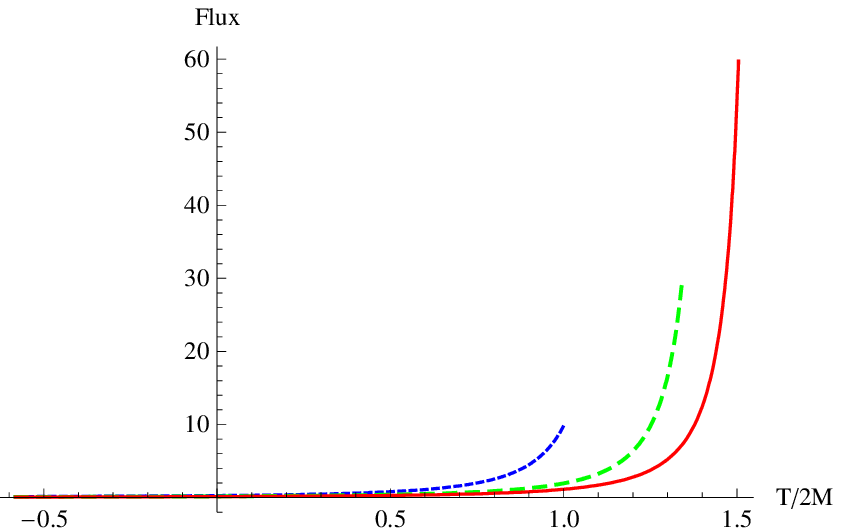}}
\caption{\small The normalized flux for the
  timelike horizon-forming trajectory with $R_0=10M$ and
  $2M\kappa=0.1$, for three $r=\,$const observers inside the
  shell, as a function of Minkowski time $T$. The curves correspond to
  : $r=1.25M$ (solid red), $r=1.5M$ (long dashed green), $r=2M$ (short
  dashed blue). The left panel shows the initial, trajectory dependent
  phase, and the right panel shows the later phase. Each trajectory is
  tracked until the shell crosses it.} 
\la{flux-ob3}
\end{figure}
Using \eqn{eq-fluxob3-1} the flux can be expressed as a function of
time $T$, for an observer with $r=\,$const. The structure of the
function \Cal{S}\ shows that the flux in the interior will remain zero
as long as the shell trajectory satisfies $V_s\propto v_s$, which is
the case until $\tau=0$. As a result, observers will see no flux as
long as they are to the left of the ingoing ray which enters at
$\tau=0$. (See the Penrose diagram \fig{penrose-hor}. This is in fact
true outside the shell as well.) This is also
confirmed by \fig{flux-ob3} which shows the flux as a function of time
for three $r=\,$const observers. We see that after an initial
complicated phase which depends on the trajectory details, the flux
becomes positive and steadily increases in magnitude until the time the
shell crosses the observer. This situation holds regardless of 
whether the shell crosses the observer before or after forming the
event horizon, although the flux at shell crossing is larger for
observers who are closer to the origin. 

\subsection{Asymptotic trajectory}
\noindent
In this section we discuss a class of trajectories first described in
\Cite{paddy-aseem}, which are timelike throughout and asymptotically
approach a final radius larger than the \Sch\ radius of the
shell. Classically the interior geometry is Minkowski spacetime and
the exterior is \Sch\ spacetime, with no event horizon. Although such
a trajectory was discussed in detail in \Cite{paddy-aseem}, there are
two reasons for returning to it here. Firstly, as mentioned earlier,
the treatment in \Cite{paddy-aseem} was mainly analytical and hence the
inversions required to calculate $G(u)$ could not be performed. This
can now be done in our current numerical setup, and we will display
results that corroborate those in \Cite{paddy-aseem}. Secondly, a
numerical analysis allows us to further explore the semiclassical
environment of this trajectory. Specifically, we will show below that
such trajectories will generically give rise to a pulse of radiation
which travels inwards, reflects at the origin and eventually reaches
an external observer at late times. Mathematically, this new feature
arises solely from the functional inversions at the shell boundary,
and was (not surprisingly) missed in \Cite{paddy-aseem}. 

The asymptotic trajectory is parametrized by Eddington retarded time
$u$ (which is a good parameter everywhere since there is no horizon),
and is described by the following prescription, borrowed from
\Cite{paddy-aseem} : 
\be
U_s^{\prime}(u) = \eps + \bigg(\left(1-\frac{1}{R_0}\right)^{1/2} - \eps 
\bigg) e^{-\alpha(u)} \equiv \eps + Ae^{-\alpha(u)}\,,  
\la{eq-traj-asymp-1}
\ee
where, as before, the prime is a derivative with respect to the
argument and we are working with the non-dimensionalised form of the
coordinates. Here $0<\eps<(1-1/R_0)^{1/2}$ is a fixed constant, 
\be
\alpha(u)=\frac{\theta(u)}{2}\int_0^u{d\ti{u} h(\ti{u})}\,,
\la{eq-traj-asymp-2}
\ee
where $\theta(u)$ is the Heaviside step function and $h(u)$ is chosen
to have the following asymptotic behaviour
\begin{align}
h(u) = 0\,, u\leq0~~&;~~ h^\prime(u)=0\,, u\leq0\,, \nonumber\\
h(u\to\infty)\to1~~&;~~ h^\prime(u\to\infty)\to0\,,
\la{eq-traj-asymp-3}
\end{align}
and we require the asymptotic values for $h$ and $h^\prime$ to be
achieved exponentially fast, with a time scale determined by $M$. For
this paper, we choose $h(u)=\theta(u)\tanh^2(\lambda u)$ for some
constant $\lambda$. This is slightly different from the choice in
\Cite{paddy-aseem}; this change does not alter the qualitative
behaviour of the trajectory and was made to improve the numerical
behaviour of various subsequent functions. The trajectory radius $R_s$
satisfies the equation
\be
2R_s^\prime(1-U_s^\prime) = U_s^{\prime2} -
\left(1-\frac{1}{R_s}\right)\,,
\la{eq-traj-asymp-4}
\ee
so that $R_s$ starts at a constant radius $R_0$, falls inwards and
asymptotically approaches a final radius $R_f\equiv 1/(1-\eps^2)$. The
Penrose diagram for such a trajectory was shown in \Cite{paddy-aseem}
(see their Fig. 3).

We again consider two classes of observers : exterior observers at
some constant radius $r>R_0$, and interior observers at some constant
radius $r<R_f$. For brevity we ignore observers who are initially
inside the shell and eventually come into the \Sch\ region, since such
observers will not see any effects that are absent for the previous
two classes. For the exterior observers, the normalized flux is easily
seen to be identical in expression to the one calculated for the
previous trajectory
\be
\Cal{\wti F} = (192\pi) F_u\left[\frac{dG}{du}\right]\,,
\la{eq-flux-asymp-ob1-1}
\ee
where
\be
G(u) = v_s(u_1(u)) ~~;~~ V_s(u_1) = U_s(u)\,,
\la{eq-flux-asymp-ob1-2}
\ee
which is familiar from earlier, with $\tau$ being replaced by $u$. The
expression for $F_u[dG/du]$ is somewhat simpler in terms of $u$,
\be
F_u\left[\frac{dG}{du}\right] = u_1^{\prime2}(u) \left.\left(
F_y[v_s^\prime(y)]\right)\right|_{y=u_1(u)} + F_u[u_1^\prime(u)]\,.
\la{eq-flux-asymp-ob1-3}
\ee
Although these expressions may appear different from Eqn. (27) of
\Cite{paddy-aseem}, the two sets are in fact equivalent. The specific
splitting used in \Cite{paddy-aseem} was conceptually easier to
tackle, while the expressions above are easier to handle numerically.
\begin{figure}[t]
\centering
\subfigure[]{
\includegraphics[width=0.45\textwidth]{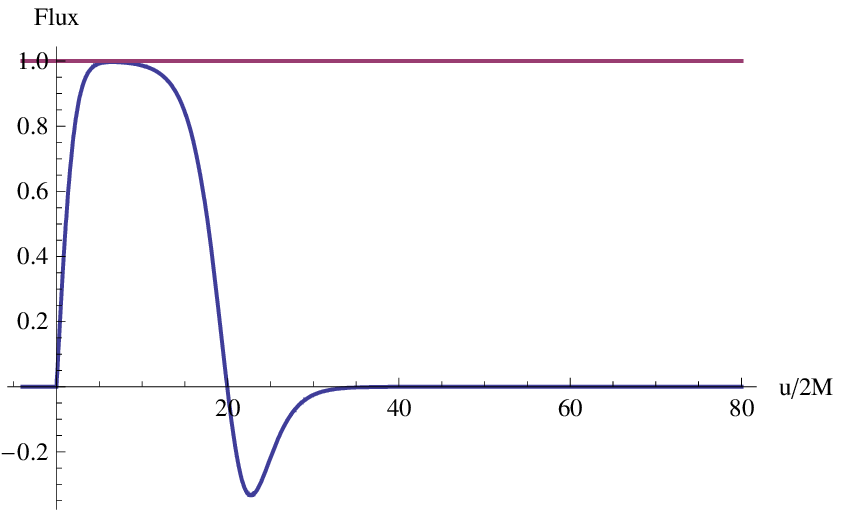}}
\subfigure[]{
\includegraphics[width=0.45\textwidth]{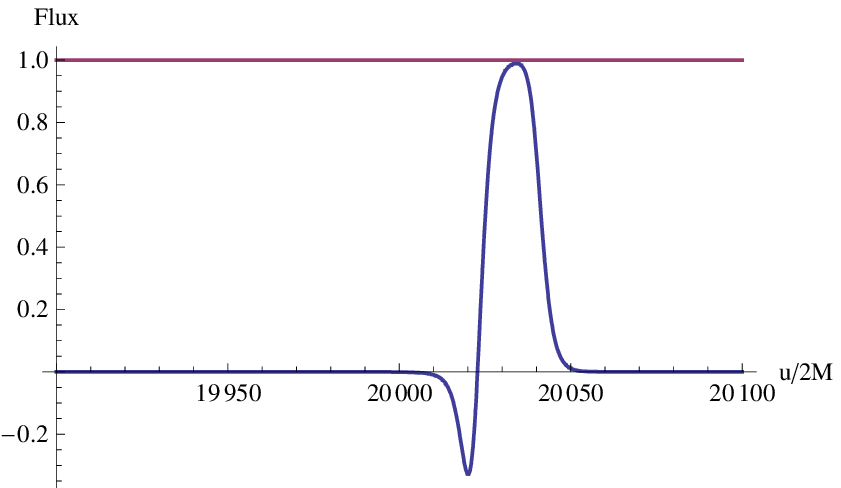}}
\caption{\small The normalized exterior flux as a function of $u$, for
  the timelike asymptotic trajectory with $R_0=3M$, $\eps=10^{-4}$ and 
  $2M\lambda=0.25$. The left panel shows the initial phase which shows
  a thermal behaviour followed by an exponential decay. The right
  panel shows the flux for the same configuration at a much later
  time. The second, ``pseudo-thermal'' phase can be traced back to a
  pulse generated at an inflection point of the trajectory, as
  discussed in the text.}
\la{flux-asymp-ob1}
\end{figure}

Since the dependence of the flux on the model parameters was discussed
in detail in \Cite{paddy-aseem}, here we will simply display results
for a choice of parameter values that enhances the effects we want to
focus on. The effects themselves are generic and do not depend on the
specific parameter choices we make. Figure \ref{flux-asymp-ob1} shows
the exterior flux as a function of $u$, for a trajectory with
$R_0=3M$, $\eps=10^{-4}$ and $2M\lambda=0.25$. As expected for such a
trajectory, we see an initial phase in the flux which closely
resembles that in the conventional trajectory : the asymptotic
trajectory initially behaves as if it were heading towards horizon
formation. Eventually however, the effects of the nonzero \eps\ are
felt and the flux exponentially decays to zero. This part is captured
in the left panel of the figure, and was also described in
\Cite{paddy-aseem}. The right panel shows the flux for the \emph{same}
trajectory, but at much later times. We see a \emph{second}
pseudo-thermal phase (more on this nomenclature later), which rises
from zero, reaches and stays at the Hawking value, and then also
exponentially decays. The origin of this second phase will become
clear soon, when we discuss the flux seen by the interior observers. 

For the interior observers, the flux calculation again proceeds
identically to that for the horizon-forming trajectory. In fact, it is
easily verified that the function \Cal{S}\ defined in
\eqn{eq-fluxob3-3} is parametrization independent, so that the
normalized interior (outgoing) flux for the present case is given by 
\begin{align}
\Cal{\wti F} &= \Cal{S}(u_1) - \Cal{S}(u_2)\,,
\la{eq-flux-asymp-ob2-1}\\
\Cal{S}(u) &= \frac{192\pi}{V_s^{\prime2}(u)}\left(
F_u[v_s^\prime(u)] - F_u[V_s^\prime(u)] \right)\,,
\la{eq-flux-asymp-ob2-2}
\end{align}
with $u_1$ and $u_2$ defined in terms of $(T,r)$ via
\be
V_s(u_1) = T-r ~~;~~ V_s(u_2) = T+r\,.
\la{eq-flux-asymp-ob2-3}
\ee
\begin{figure}[t]
\centering
\subfigure[]{
\includegraphics[width=0.45\textwidth]{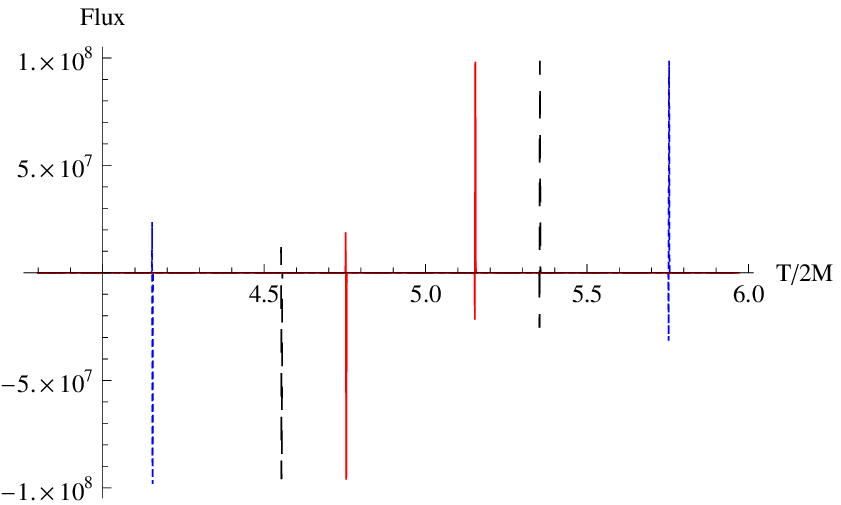}}
\subfigure[]{
\includegraphics[width=0.45\textwidth]{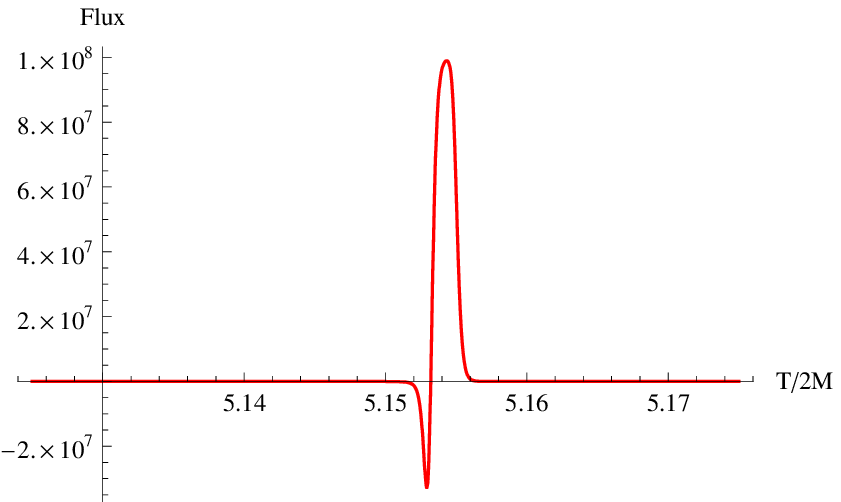}}
\caption{\small The normalized interior flux as a function of
  Minkowski time $T$, for the timelike asymptotic trajectory with
  $R_0=3M$, $\eps=10^{-4}$ and $2M\lambda=0.25$. The left panel shows
  the flux seen by three observers at constant $r$ : $r=0.4M$ (solid
  red), $r=0.8M$ (long dashed black) and $r=1.6M$ (short dashed
  blue). The right panel shows details of the pulse (shown on the
  outgoing leg for the observer with $r=0.4M$).}
\la{flux-asymp-ob2}
\end{figure}
As before, the flux will be nonzero only on the right of the ingoing
ray that enters at $u=0$, since in the region to the left of this ray
we have $V_s\propto v_s$. Figure \ref{flux-asymp-ob2} shows the flux
seen by three observers at three different constant radii $r$, for the
same configuration as in \fig{flux-asymp-ob1}. Each observer sees a
strong incoming pulse of radiation and an outgoing pulse at a later
time. By studying the time intervals between the pulses seen by
various observers, it is possible to show that there is only one pulse
which starts at the shell at a certain time, travels inwards and gets 
reflected at the origin, and is then seen by the observers again on
its way out. This is also consistent with the fact that the second
pulse seen by any interior observer is the precise mirror image of the
first (although this is not obvious from the figure). The right
panel in \fig{flux-asymp-ob2} shows the detailed structure of the
pulse. 
\begin{figure}[t]
\centering
\subfigure[]{
\includegraphics[width=0.425\textwidth]{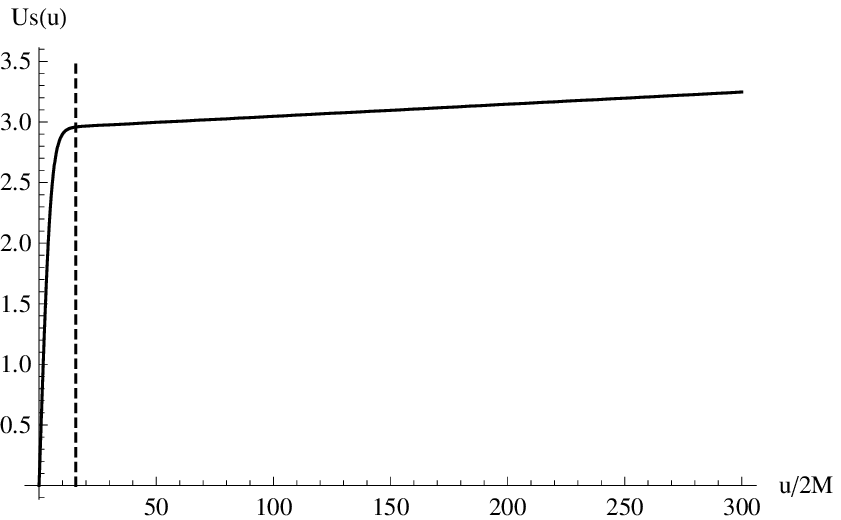}}
\subfigure[]{
\includegraphics[width=0.425\textwidth]{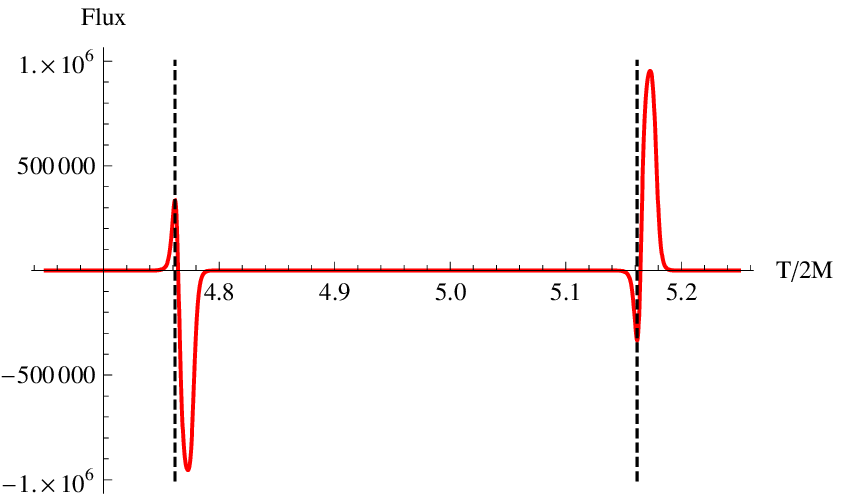}}\\
\subfigure[]{
\includegraphics[width=0.425\textwidth]{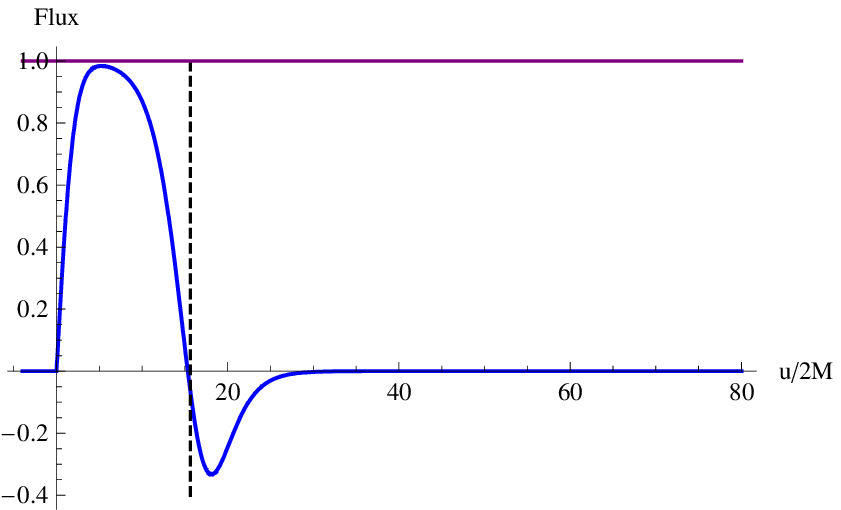}}
\subfigure[]{
\includegraphics[width=0.425\textwidth]{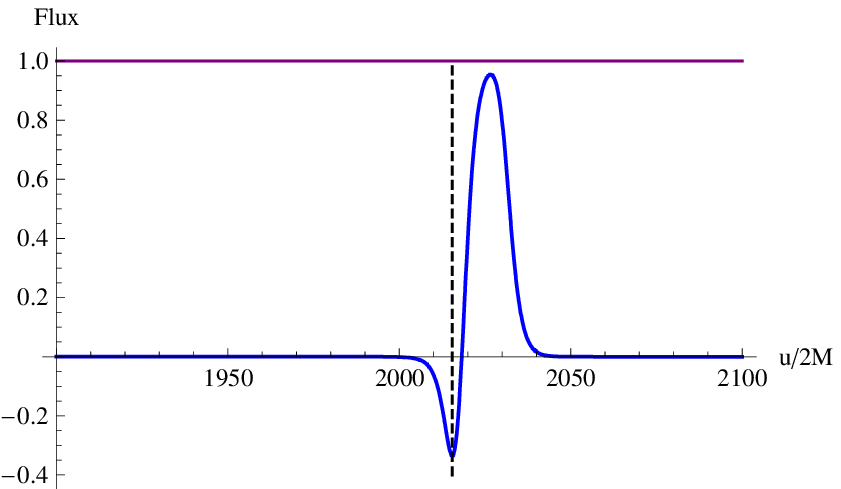}}
\caption{\small The behaviour of the pulse which arises in the
  interval when $U_s$ changes its behaviour (panel (a)), for the
  asymptotic trajectory configuration $R_0=3M$, $\eps=10^{-3}$ and
  $2M\lambda=0.25$. The vertical dashed line in panel (a) marks
  $u=u_{\rm change}$, and the vertical lines in the other panels track
  the ray \Cal{R}\ that enters the shell at $u=u_{\rm change}$. Panel
  (b) : the interior normalized flux for the observer at $r=0.4M$. Vertical
  lines mark the times when this observer meets the ray \Cal{R},
  before and after reflection. Panel (c) : The initial phase of the
  exterior flux. Vertical line marks the ray \Cal{R}\ on the way
  in. Panel (d) : The late phase of the exterior flux. Vertical line
  marks the ray \Cal{R}\ on the way out.}
\la{flux-asymp-pulse}
\end{figure}

This discussion also suggests that this pulse
should be detected by \emph{exterior} observers as well, and in fact
one finds that the second pseudo-thermal phase referred to
earlier, is precisely the remnant of this pulse after it exits the
shell. Further, the origin of this pulse can be traced to the time
interval in which the behaviour of $U_s$ changes from $U_s \sim
(1-1/R_0)^{1/2}u$ to $U_s\sim\eps u$ (see \eqn{eq-traj-asymp-1}). To
visually demonstrate this we define the time $u_{\rm change}$ such
that the slope $U_s^\prime(u_{\rm change})=e\eps$, i.e. one
$e$-folding away from its final asymptotic value. In
\fig{flux-asymp-pulse}, the 
vertical dashed line keeps track of the ray which enters at $u_{\rm
  change}$. This figure uses the same trajectory configuration 
as in \figs{flux-asymp-ob1} and \ref{flux-asymp-ob2} except for the
value of \eps, which is now $\eps=10^{-3}$. Figure
\ref{flux-asymp-pulse} when compared with \figs{flux-asymp-ob1} and
\ref{flux-asymp-ob2} also suggests that the pulse height in the
interior scales approximately as $\eps^{-2}$, which also follows from
considering the structure of \Cal{S}, and that the time at which the
second phase occurs in the exterior, scales roughly as
$\eps^{-1}$. This is also supported by a more detailed numerical
analysis (not shown). 

One significant aspect of the pulse of radiation is that it is very
sharply peaked in the interior of the shell, and becomes diffuse once
it exits to the exterior. In fact one can check that the height of the
pulse in the exterior never exceeds the Hawking value, instead the
width increases with decreasing \eps, and the flux in the exterior
pulse spends a 
longer time at the Hawking value. This is why we referred to this
phase as ``pseudo-thermal'' earlier. It will be very interesting to
see the signatures left by this pulse in the spectrum of particles in
various regions. This however requires a calculation of the Bogolubov
coefficients, which is complicated by the presence of numerical
inversions, and is beyond the scope of this work. Another tantalizing
prospect is that such a pulse could possibly lead to astrophysical
signatures in objects which undergo phases of halted collapse without
forming a horizon. This would, however, require some kind of
amplification mechanism, since the total amount of gravitational
energy released in the pulse appears to be very small (while the
pulse height can become large, the pulse \emph{width} decreases with
decreasing \eps).  

\section{Discussion}
\la{discuss}
\noindent
We have explored in detail the semiclassical environment of a
collapsing shell of matter. Our main objective was to study the
semiclassical flux observed by a variety of observers, both inside and
outside the shell. We have studied two types of trajectories : those
which form a horizon and those which do not. 

For the horizon forming trajectories, we reproduced the well known
result of an asymptotically constant flux seen by observers outside
the horizon. Additionally we demonstrated that freely falling
observers measure an \emph{outgoing} flux which generically increases
without bound as they approach the singularity. Our calculation
assumed zero backreaction, and it is possible that the infinity we see
in our flux can somehow be cured  if backreaction were correctly
accounted for throughout. It is hard to guess how exactly this might
happen, since a self-consistent calculation of a black hole geometry
in the presence of backreaction at late stages of evaporation, does
not exist. It is possible that the resolution of this problem is
linked to the more general problem of the singularity itself.

We also saw, at least in the timelike case, that inertial observers 
\emph{inside} the shell also see a nonzero outgoing flux, which
steadily increases until the time they exit the shell. In the context
of backreaction, this has consequences for all arguments built on the
assumption that the interior geometry of the shell remains that of
Minkowski spacetime throughout. This assumption was made for
example in \Cite{paddy-aseem} while arguing that the semiclassical
backreaction can at best delay the formation of the event horizon by a
short amount of time. While it is unlikely that including a flux of
the same order of magnitude as the Hawking flux would overturn the
conclusions of the arguments in \Cite{paddy-aseem}, it is nevertheless
important to rigorously account for this additional flux and build a
complete picture. Specifically, it is interesting to ask what the
self-consistent geometry in the interior will be, after accounting for
backreaction. We will return to this question in future work.

In the case of the asymptotic trajectory which halts collapse before
horizon formation, we corroborated the analysis of \Cite{paddy-aseem}
by a full numerical calculation, and extended the results to include
the flux seen by interior observers. In the process we uncovered an
interesting, generic feature of such trajectories, in the form of a
pulse of radiation which is sharply peaked in the interior but appears
to thermalize as it emerges from the shell. It remains to be seen
whether this feature might lead to astrophysically interesting
signatures. 

For both types of trajectories discussed here, it will be interesting
to determine the nature of the particle spectrum, for which we need to
calculate the Bogolubov coefficients in the respective
geometries. This is work in progress. We expect our current results to
be useful in addressing the broader issues mentioned in the
Introduction, specifically questions related to the interpretation of
thermodynamic variables such as entropy, for a variety of classes of
observers, and in studying the role and nature of the semiclassical
backreaction.

\acknowledgments
We wish to thank T. Padmanabhan for
suggesting this project, and for many insightful discussions. AP
thanks Dawood Kothawala for useful discussions, and gratefully
acknowledges hospitality at IUCAA where most of this work was
completed.

\end{document}